\documentclass[twocolumn,preprintnumbers,amsmath,amssymb]{revtex4}

\usepackage{graphicx}% Include figure files
\usepackage{dcolumn}% Align table columns on decimal point
\usepackage{bm}% bold math
\begin{document}

\preprint{APS/123-QED}

\title{On the missing link between pressure drop, viscous dissipation, and the turbulent energy spectrum}% Force line breaks with \\

\author{Arnoldo Badillo and Omar K. Matar}
% \altaffiliation[Also at ]{aaa Department, XYZ University.}%Lines break automatically or can be forced with \\
 \email{o.matar@imperial.ac.uk}
\affiliation{%
Department of Chemical Engineering,\\
Imperial College London, Prince Consort Road, SW7 2AZ, UK.
}%

\date{\today}% It is always \today, today,

\newcommand{\Rey}{\mbox{\textit{Re}}}   % Reynolds number
\newcommand{\vel}{\textbf{u}}
\newcommand{\stress}{\stackrel{\leftrightarrow}{\boldsymbol{\sigma}}}
\newcommand{\corr}{\stackrel{\leftrightarrow}{\textbf{R}}}
\newcommand{\fourier}{\hat{\textbf{R}}}
\newcommand{\deviatoric}{\stackrel{\leftrightarrow}{\boldsymbol{\tau}}}
\newcommand{\identity}{\stackrel{\leftrightarrow}{\textbf{I}}}

\newcounter{defcounter}
\setcounter{defcounter}{0}
\newenvironment{myequation}{%
\addtocounter{equation}{-1}
\refstepcounter{defcounter}
\renewcommand\theequation{A\thedefcounter}
\begin{equation}}
{\end{equation}}

\begin{abstract}
After decades of experimental, theoretical, and numerical research in fluid dynamics, many aspects of turbulence remain poorly understood. The main reason for this is often attributed to the multiscale nature of turbulent flows, which poses a formidable challenge. There are, however, properties of these flows whose roles and inter-connections have never been clarified fully. In this article, we present a new connection between the pressure drop, viscous dissipation, and the turbulent energy spectrum, which, to the best of our knowledge, has never been established prior to our work. We use this finding to show analytically that viscous dissipation in laminar pipe flows cannot increase the temperature of the fluid, and to also reproduce qualitatively Nikuradse's experimental results involving pressure drops in turbulent flows in rough pipes.

\end{abstract}

%\pacs{}
\maketitle

%%%%%%%%%%%%%%%%%%%%%%%%%%%%%%%%%%%%%%%%%%%%%%%%%%%%%%%%%%%
%                INTRODUCTION                             %
%%%%%%%%%%%%%%%%%%%%%%%%%%%%%%%%%%%%%%%%%%%%%%%%%%%%%%%%%%%
%\section{Introduction}
For centuries, turbulence has remained stubbornly one of the most challenging problems to solve in classical physics. The multiscale character of turbulent flows, and the poorly understood fundamental mechanisms for distributing the kinetic energy among the large number of harmonics which constitute the fluctuating velocity field, are among the main difficulties that complicate significantly the analysis of turbulent flows. It is well accepted that vortex stretching is one of the main mechanisms responsible for transferring energy from low to high wave numbers. This mechanism, however, only exists in viscous, three-dimensional turbulence, and has no analogue in quantum \cite{Barenghi}, as well as two-dimensional turbulence \cite{Tran}. The pictorial energy cascade proposed by Richardson \cite{Richardson}, and then recovered theoretically first by Kolmogorov \cite{Kolmogorov1, Kolmogorov2}, and then, independently, by Onsager \cite{Eyink}, does not reveal much about the energy transfer between consecutive harmonics; it only represents an `equilibrium' state of energy distribution arising from a balance between the fluxes of energy going up and down the cascade. Despite the fact that viscous dissipation is irrelevant for quantum turbulence, the scaling laws dominating its energy cascade are the same as those in viscous, three-dimensional turbulence \cite{Maurer,Maltrud,Smith}. What, then, is the true role of viscous dissipation in the formation of the energy cascade, and what is the actual physical meaning of `dissipation' and its connection to macroscopic quantities such as the pressure drop?

We start our discussion with the enthalpy equation for a pure single phase incompressible flow, \mbox{$\rho D_{t} h +\nabla\cdot\textbf{q}=  D_{t} P+\deviatoric\colon\nabla\textbf{u}$} (see Appendix for a detailed derivation), where $D_t=\partial_t + \textbf{u}\cdot\nabla$ is the material derivative, $h$ the enthalpy per unit mass, $\textbf{q}$ the heat flux, $P$ the pressure, and $\deviatoric$ the viscous stress tensor. It can be shown (see Appendix) that for a fully-developed laminar, and adiabatic flow in a cylindrical pipe, the integration of the enthalpy equation in the whole domain $\Omega$, leads to a pressure drop that balances exactly the total viscous dissipation:
\begin{equation}
\Delta P = -\frac{1}{VA}\int_{\Omega}\deviatoric\colon\nabla\textbf{u}d\Omega,
\label{eqn:pressure_drop_1}
\end{equation}
where $V$ is the mean flow velocity, and $A$ the pipe cross-sectional area. This linear relationship between the total dissipation, given by the volumetric integral in Eq. (\ref{eqn:pressure_drop_1}), and the external forcing, corresponding to the pressure drop, is one of the central results of the present study. Interestingly, a linear relationship between dissipation and external stimuli has recently been observed in turbulent quantum gases (see Fig. 2e in Ref. \cite{Navon}). The validity of Eq. (\ref{eqn:pressure_drop_1}) implies that for incompressible, fully-developed pipe flow, a fluid cannot increase its temperature due to viscous dissipation. This appears to be a somewhat counterintuitive conclusion because the energy per unit volume injected into the fluid, in the form of pressure at the pipe inlet, must be dissipated into some form of microscopic energy. In fact, a detailed derivation of the entropy equation (see Appendix) shows that the entropy of the fluid does change with the viscous dissipation, that is, \mbox{$\rho T D_t s = -\nabla\cdot\textbf{q}+\deviatoric\colon\nabla\textbf{u}$}.

A change in the entropy of the system indicates that viscous dissipation is indeed a form of heat, but not necessarily of the sensible type. From a statistical mechanics view point, the entropy, $S$, defined in terms of microstates probabilities $\pi_i$ as $S = -k_B \sum^N_{i=1}{\pi_i \ln\pi_i}$, is a measure of the total number of accessible microstates to a system, $N$, under a given pressure and temperature. By changing the internal energy of a microstate, we alter its probability and, therefore, the entropy of the system. By applying a higher pressure at the inlet of a pipe, we restrict molecules from sampling certain configurations, which at lower pressures, are readily available. Thus, when the pressure drops along the pipe, the entropy increases by having a larger number of accessible molecular configurations. When we perform isothermal (i.e. neglecting the energy equation) incompressible direct numerical simulations of turbulent flows, we are implicitly satisfying Eq. (\ref{eqn:pressure_drop_1}). A different scenario exists for compressible flows, where fluid compression leads to an enthalpy change.

To investigate the relevance of Eq. (\ref{eqn:pressure_drop_1}) and its connection to the energy spectrum, we follow the work of Gioia and Chakraborty \cite{Gioia1}. They investigated the relationship between the friction factor, $f = 8\tau_w/\rho V^2$ (expressed in terms of the wall shear stress $\tau_w$), and the phenomenological turbulent energy spectrum. Considering the Darcy-Weisbach equation, we can express the time-averaged (denoted by angular brackets) pressure drop as
\begin{equation}
	\left\langle \Delta P\right\rangle = -f \frac{L}{D_H} \frac{\rho V^2}{2},
	\label{eqn:darcy-weisbach}
\end{equation}
where $D_H$ the hydraulic pipe diameter, $\rho$ the fluid density and $L$ the pipe length. The negative sign is to enforce a decrease in the pressure along the pipe. By taking the time average of Eq. (\ref{eqn:pressure_drop_1}) and substituting it into Eq. (\ref{eqn:darcy-weisbach}), we can express the friction factor in terms of the viscous dissipation, instead of the wall shear stress:
\begin{equation}
f = \frac{64}{\Rey}\frac{1}{8\pi L V^2}\left(
\int_{\Omega}\frac{\overline{\epsilon}_0}{\nu}d\Omega
+\int_{\Omega}\frac{\epsilon'_0}{\nu}d\Omega
\right)
\label{eqn:friction}
\end{equation}
where $\nu$ is the kinematic viscosity, and \mbox{$\rho^{-1}\left\langle \deviatoric\colon\nabla\textbf{u}\right\rangle = \overline{\epsilon_0} + \epsilon_0'$} is defined as the total time-averaged dissipation with $\overline{\epsilon_0}/\nu = \nabla\overline{\textbf{u}}\colon\nabla\overline{\textbf{u}}+\nabla\overline{\textbf{u}}^T\colon\nabla\overline{\textbf{u}}$ and $\epsilon_0'/\nu = \left\langle \nabla\textbf{u}'\colon\nabla\textbf{u}' + \nabla\textbf{u}'^T\colon\nabla\textbf{u}'\right\rangle$. Here, we have used the Reynolds decomposition for the velocity field $\textbf{u}\left(\textbf{x},t\right)=\overline{\textbf{u}}\left(\textbf{x}\right)+\textbf{u}'\left(\textbf{x},t\right)$. The Reynolds number, $\Rey = V D_H/\nu$, is defined in terms of the hydraulic pipe diameter, $D_H$.

%Equation (\ref{eqn:friction}) represents a substantial difference from the friction factor expressed in terms of the wall shear stress. For fully-developed laminar pipe flow, the friction factor calculated from the wall shear stress or Eq. (\ref{eqn:friction}) leads to the well known relation \mbox{$f = 64/\Rey$}. However, for turbulent flows, the physical interpretation of both equations is quite different.

The main difficulty faced by Gioia and Chakraborty \cite{Gioia1}, was the determination of the wall shear stress, and how to relate it to the turbulent energy spectrum and the surface roughness. They postulated that the wall shear stress %which is formally defined for unidirectional flows as $\tau_w=\mu\left(\partial u/\partial y\right)_w$, with $u$ the axial velocity and $y$ the radial coordinate,
can be expressed in terms of a momentum transfer between the bulk flow and the flow near the wall, characterized by a velocity scale $u_s$. Expressing the wall stress as $\tau_w=\rho V u_s$, the problem reduces to finding the velocity scale, which might also vary with the amplitude of the surface roughness. Furthermore, to determine the velocity scale they assumed the validity of the phenomenological energy spectrum in the proximity of the wall, which is questionable. It is well known, however, that to derive the phenomenological spectrum, the turbulent structures must be rotationally- and translationally-invariant; that is, the flow is isotropic and homogenous, which is not the case near the wall. Aware of this limitation, Gioia and Chakraborty \cite{Gioia1} argue that even for anisotropic and inhomogeneous flows, the phenomenological theory still represents a good approximation. Thus, by using directly the energy spectrum $E\left(q\right)$ (including corrections for the energetic and dissipative ranges), they obtain the velocity scale as $u_s^2 = \int_{1/s}^{\infty}{ E\left(q\right)}dq$. Here, the integration is carried out for wave numbers higher than $1/s = 1/\left(r+a\eta\right)$ (where $s$ is not to be confused with the entropy), where $a$ is a dimensionless constant, $r$ the size of the surface roughness, and $\eta$ the Kolmogorov length scale. Hence, they are effectively filtering out all eddies larger than $s$ in the calculation of the velocity scale.

It is remarkable that for $\Rey \rightarrow \infty$, the friction factor in the work of Gioia and Chakraborty \cite{Gioia1} scales as $f \sim (r/R)^{1/3}$ despite the absence of viscous dissipation correction: without this, the total dissipation diverges when using the Kolmogorov spectrum. The authors recovered the proper scaling for the friction factor, observed in Nikuradse's experimental results, only by selecting an appropriate cutoff wave length ($1/s$) in the integration of the energy spectrum. The lack of physical meaning of a constant friction factor, in the absence of a correction in the dissipation range, is what motivated us to investigate a different approach to explain Nikuradse's findings.

In our analysis, we start by investigating the direct connection between the friction factor, defined in Eq. (\ref{eqn:friction}) with the turbulent energy spectrum. We now introduce the tensor correlation $\corr_{ijk}\left(\textbf{x},\delta\textbf{x}_k\right)=\left\langle u'_i\left(\textbf{x},t\right) u'_j\left(\textbf{x}+\delta\textbf{x}_k,t\right) \right\rangle$, where $i$ and $j$ represent the components of the fluctuating velocity field and $k$ the direction in which the correlation is being calculated. Since $\delta\textbf{x}_k$ can be chosen independently from $\textbf{x}$, $\corr_{ijk}$ is a 3-rank tensor, with eighteen independent components. The position and orientation-dependent energy spectrum is directly obtained by applying the Fourier transform to the tensor correlation, this is, \mbox{$\fourier_{ijk}\left(\textbf{x},\textbf{q}_k\right) = \int_{-\infty}^{\infty} e^{-2\pi i\textbf{q}_k\cdot\delta\textbf{x}_k }\corr_{ijk}\left(\textbf{x},\delta\textbf{x}_k\right) d\delta\textbf{x}_k$}. Since no assumption has been made in the definition of the tensor correlation, this definition of the energy spectrum is exact. Using a Taylor expansion to express the tensor correlation in terms of a position and orientation-independent correlation plus a corrections, we have $\corr_{ijk}=\corr_0+\left.\partial_{\textbf{x}}\corr_{ijk}\right|_0 \Delta\textbf{x}+\left.\partial_{\delta\textbf{x}}\corr_{ijk}\right|_0 \Delta\delta\textbf{x}_k+\cdots$. Here, $\corr_0$ is the homogeneous tensor correlation under rotational invariance, which has six independent components. We assume now that the Taylor expansion is convergent and that first-order corrections are the leading terms. The first and second terms represent the departure from homogeneity and isotropy, respectively. Thus, the homogeneous isotropic tensor $\corr_0$ can be interpreted as a spherical average followed by space average of $\corr_{ijk}$. With this, the energy spectrum can be finally written as $\fourier_{ijk}\left(\textbf{x},\textbf{q}_k\right) = \hat{R}_0\left(q\right)+\delta\fourier_t\left(\textbf{x},q\right)\Delta\textbf{x}+\delta\fourier_r\left(\delta\textbf{x},\delta\textbf{q}_k\right)\Delta\delta\textbf{x}+\cdots$. This enables us to calculate each component of the viscous dissipation easily, that is, $\left\langle\left(\partial_i u'_j\right)^2\right\rangle = \int_{-\infty}^{\infty}\textbf{q}_i^2\fourier_{jji}\left(\textbf{x},\delta\textbf{x}_i=0\right)d\textbf{q}_i$ and $\left\langle\partial_i u'_j\partial_j u'_i \right\rangle = \int_{-\infty}^{\infty}\textbf{q}_{\left(ij\right)}^2\fourier_{ji\left(ij\right)}\left(\textbf{x},\delta\textbf{x}_{\left(ij\right)}=0\right)d\textbf{q}_{\left(ij\right)}$. These expressions are exact and the subscripts $\left(ij\right)$ indicate a spectrum obtained along $\delta\textbf{x}_k=\delta\textbf{x}_i+\delta\textbf{x}_j$. Hence, the connection between the energy spectrum and the pressure drop is directly realized trough Eqs. (\ref{eqn:darcy-weisbach}-\ref{eqn:friction}), without the need of any {\it ad hoc} wavelength cutoff in the integral. Following Taylor's \cite{Taylor} identities for isotropic turbulence, we have that the fluctuating component of the dissipation rate can be approximated by \mbox{$\epsilon'_0\approx 30\nu\int_0^\infty q^2 \hat{R}_0 dq + \mathcal{O}\left(\Delta\textbf{x}+\Delta\delta\textbf{x}_k\right)$}.

We now analyze the asymptotic behavior of Eq. (\ref{eqn:friction}). For $\Rey\rightarrow 0$, $\int_{\Omega}\overline{\epsilon}_0/\nu d\Omega\rightarrow 8\pi L V^2 $ and $\epsilon'_0 \rightarrow 0$. Thus, the appropriate scaling for laminar flow $f\sim\Rey^{-1} $ is recovered directly. Conversely, for $\Rey\rightarrow \infty$, $\int_{\Omega}\overline{\epsilon}_0/\nu d\Omega\rightarrow 0 $ and $\int_{\Omega}\epsilon'_0/\nu d\Omega\sim L V^2 \Rey$, which is a result obtained from imposing a constant friction factor for high Reynolds numbers. Since our hypothesis is that only viscous dissipation plays a role in the friction factor (or pressure drop), the important question is to know whether the two scalings, observed in rough pipe flows, $f\sim\Rey^{-1/4}$ and $f\sim \left(r/R\right)^{1/3}$ as $\Rey \rightarrow \infty$, can be explained solely in terms of the fluctuating component of the viscous dissipation. We, therefore, center our attention on the second integral in Eq. (\ref{eqn:friction}). Neglecting the correction due to anisotropy and inhomogeneity in the fluctuating part of the viscous dissipation, this integral reduces to $\int_{\Omega}\epsilon'_0/\nu d\Omega \approx 30 \Omega \int_0^\infty q^2 \hat{R}_0 dq$, with $\Omega$ the pipe volume. Because $\hat{R}_0$ is isotropic and homogeneous, we will also make use of the phenomenological spectrum to determine this integral. Since we are interested in the dissipative range of the spectrum, we consider a correction $C_d\left(q\right)$ only for this regime. A correction for the inertial regime like the one used by Gioia and Chakraborty \cite{Gioia1}, is only needed to obtain a finite integral length scale defined as $l_{int}=\int_0^\infty q^{-1}\hat{R}_0 dq/\int_0^\infty \hat{R}_0 dq$. Without the correction for the inertial regime, the integral length scale diverges. Hence, we choose $\hat{R}_0 = A\tilde{\epsilon}^{2/3}q^{-5/3}C_d\left(q\right)$, with $A$ a dimensionless constant and $\tilde{\epsilon}$ a dissipation rate scale (which must not be confused with $\epsilon'_0$). Thus, $\int_{\Omega}\epsilon'_0/\nu d\Omega \approx 30\Omega A\tilde{\epsilon}^{2/3}\int_0^\infty q^{1/3} C_d\left(q\right) dq$. It is clear from this result that without the correction for the dissipative regime, the total dissipation diverges.

To recover the proper scaling for the turbulent friction factor, the problem reduces to choosing an appropriate correction $C_d(q)$. For instance, we can recover the correct scaling by selecting $C_d(q)=\exp\left(-\beta \psi\left[R/\eta,\Rey, r/R\right] \eta q \right)$, with $\psi\left[R/\eta,\Rey, r/R\right]$ a dimensionless function, $\eta = \nu^{3/4}\tilde{\epsilon}^{-1/4}$ the Kolmogorov's length scale, and $\beta=\left(60A I_0 C_{\epsilon}^{11/12}\right)^{3/4}$ a dimensionless constant. By choosing \mbox{$\psi=\left(2R/ \eta\right)^{1/4}\left(A_2\left(1-\phi_2\right)+A_3\phi_2\left(\Rey^{3/4}\left(r/R\right)\right)^{1/3}\right)^{-3/4}$} (see Appendix) we recover the scaling $f_T\sim\Rey^{-1/4}$ and $f_T\sim \left(r/R\right)^{1/3}$. The blending function $\phi_2=1-\left[1+\left(z/z_0\right)^2\right]^{-1}$, with $z=\Rey^{3/4}\left(r/R\right)$, bridges the Blasius and Strickler regimes. The constants $z_0=35.5$, $A_2 = 0.32 $, $A_3=0.147$ were determined from a fit to Nikuradse's data. Goldenfeld\cite{Nigel} postulated that the turbulent friction factor can be expressed in general as $f_T=\Rey^{-1/4}G\left(\Rey^{3/4}\left(r/R\right)\right)$ for the Blasius and Strickler regimes, with $G$ an unknown function of the Reynolds number and the relative surface roughness $r/R$. By plotting Nikuradse's data as $f_T\Rey^{1/4}$ vs $\left(r/R\right)\Rey^{3/4}$, Goldenfeld was able to collapse almost perfectly the experimental points onto a single curve. By utilizing a second function $\phi_1 = 0.5\left(1 + \tanh\left[\left(\Rey-\Rey_T\right)/\alpha_1\right] \right)$ to bridge the laminar and turbulent regimes, with $\Rey_T = 2944.27$ and $\alpha_1=1089.43$ (also determined from a fit), we can reproduce Goldenfeld's findings over the entire range of $\Rey$ including the laminar regime. Thus, by expressing the friction factor as $f = f_L \left(1-\phi_1\right) + f_T \phi_1$, we can closely match Nikurade's results, as demonstrated in Fig. 1.
\begin{figure}[htbp]
	\centering
		\includegraphics[width=8.5cm]{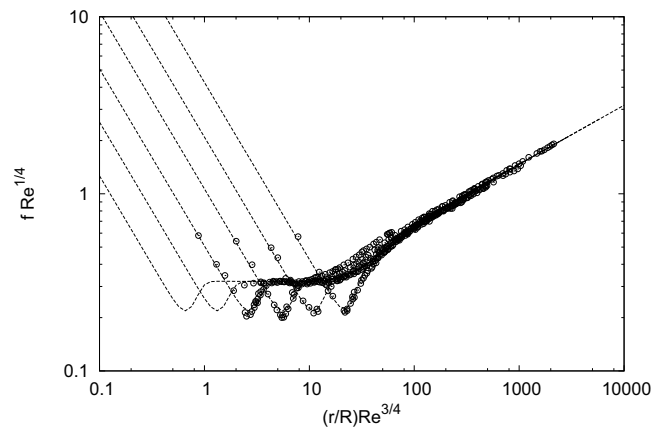}
	\label{fig:image}
	\caption{Reproduction of Nikuradse's data over the entire range of Reynolds number.}
\end{figure}

Elaborate procedures to fit Nikuradse's data have already been presented \cite{Bobby1,Bobby2}; obtaining an optimal fit to these experimental data is not our purpose. We have used Nikuradse's data only to show that when the Kolmogorov scale in the correcting factor for the dissipative range is re-scaled properly to account for surface roughness effects, the appropriate scaling for the friction factor is recovered. We also note the work of Mejia-Alvarez and Christensen \cite{Mejia}, who studied experimentally the effects of surface roughness on the velocity fluctuations near a wall, showed that the intensity of the velocity fluctuations is significantly higher for rough surfaces relative to smooth ones. Their results support our findings in two ways: first, larger fluctuations indicate higher dissipation in agreement with a higher pressure drop. Second, turbulent structures developed at or near the rough wall, penetrate into the main flow up to a certain depth. Therefore, in this wall-affected region, the dissipation scale (and therefore the Kolmogorov scale) must be modified to account for the stronger energy sinks.

In conclusion, we have uncovered a new connection between the pressure drop and the turbulent energy spectrum for incompressible flows, which necessarily implies that the energy of the fluid (per unit volume) stored in the form of pressure, is not dissipated in the form of sensible heat. For fully-developed turbulent flows, the turbulent kinetic energy is translationally-invariant and, therefore, does not decay along the pipe. This is only possible because the pressure, not the kinetic energy, is the quantity that is being dissipated into heat. We tested the new relationship between pressure drop and the turbulence energy spectrum, by introducing a variable correction to the phenomenological spectrum in the dissipative regime. Our results indicate that the Kolmogorov scale must be modified in the presence of surface roughness, to properly account for the penetration of turbulent structures, developed near rough walls, into the main stream, and the consequent higher viscous dissipation. These new findings are currently being used to unveil a new relationship between the eddy viscosity with the Reynolds number, which is expected to shed new light on the `closure' problem. \\

%Integrating the total dissipation over the volume, leads to $f=\Rey^{-1}2 D_H^2/V^2 \left(\overline\epsilon_0 + 30\int_0^\infty q^2 \hat{R}_0 %dq + \mathcal{O}\left(\Delta\textbf{x}^2+\Delta\delta\textbf{x}^2_k\right)\right)$. Due to symmetry reasons, when performing the volumetric %integral of the error, first order corrections cancel out and only second order corrections remain.

%\section*{ACKNOWLEDGEMENTS}
We acknowledge financial support from the Engineering and Physical Sciences Research Council, UK, through the Programme Grant MEMPHIS (grant number EP/K003976/1).
%
%
% 	Appendix - Conservation laws
%
%
\section*{Appendix}
\section {Conservation laws}
We start our derivation by the most general statement of conservation of a $(n)-$rank tensor $\Phi^{\left(n\right)}$ inside a control a dynamic volume $\Omega\left(t\right)$
\begin{myequation}
\frac{d}{dt}\left(\int_{\Omega\left(t\right)}\boldsymbol{\Phi}^{\left(n\right)} d \Omega\right)
+\int_{d\boldsymbol{\Gamma}\left(t\right)}\textbf{j}_{\Phi}^{\left(n+1\right)}\cdot d\boldsymbol{\Gamma}
= \int_{\Omega\left(t\right)}\textbf{S}_{\Phi}^{\left(n\right)} d\Omega
\label{eqn:conservation1}
\end{myequation}
where $\textbf{j}_{\Phi}^{\left(n+1\right)}$ represent the $\left(n+1\right)-$rank tensor flux of $\boldsymbol{\Phi}^{\left(n\right)}$ through the surface of the control volume $\boldsymbol{\Gamma}\left(t\right)$, and $\textbf{S}_{\Phi}^{\left(n\right)}$ is any source of $\boldsymbol{\Phi}^{\left(n\right)}$. Applying the divergence and Reynolds transport theorem, we arrive at the differential form of the conservation law
\begin{myequation}
\frac{\partial \boldsymbol{\Phi}^{\left(n\right)}}{\partial t}
+\nabla\cdot\left(\textbf{j}_{\Phi}^{\left(n+1\right)}-\textbf{u}_{\Gamma}\boldsymbol{\Phi}^{\left(n\right)}\right)
= \textbf{S}_{\Phi}^{\left(n\right)}
\label{eqn:conservation2}
\end{myequation}
with $\textbf{u}_{\Gamma}$ the boundary velocity of the control volume. All conservation laws can be written in this way, provided the proper expressions for total flux $\textbf{j}_{\Phi}^{\left(n\right)}$. In this work, we are concerned with mass, momentum and energy conservation, thus, the fluxes for these quantities are given by: mass $\textbf{j}^{(2)}_{\rho}=\textbf{u}\rho + \textbf{j}_d$, momentum $\textbf{j}^{(3)}_{\textbf{p}}=\textbf{u}\textbf{p} + \stress -\mu\rho\identity$, energy \mbox{$\textbf{j}^{(2)}_e=\textbf{p} e + \textbf{q} + \left(\stress-\mu\rho\identity\right)\cdot\textbf{u}$}. The term $\stress -\mu\rho\identity$ corresponds to the momentum flux due to stresses and chemical species, thus, $\left(\stress -\mu\rho\identity\right)\cdot\textbf{u}$ represents the work rate done by the fluid. By further considering external forces as a source of momentum and energy (if these forces exert work in the system), we arrive at the final set of conservation laws,
\begin{myequation}
\frac{\partial \rho}{\partial t}
+\nabla\cdot\left(\left[\textbf{u}-\textbf{u}_{\Gamma}\right]\rho\right) = -\nabla\cdot\textbf{j}_d
\label{eqn:mass}
\end{myequation}
\begin{myequation}
\frac{\partial \textbf{p} }{\partial t}
+\nabla\cdot\left(\left[\textbf{u}-\textbf{u}_{\Gamma}\right]\textbf{p}\right) =
-\nabla\cdot\left(\stress-\mu\rho\identity\right) + \Sigma\textbf{f}_{ext}
\label{eqn:momentum}
\end{myequation}
\begin{myequation}
\begin{split}
\frac{\partial \left(\rho e\right) }{\partial t}
+\nabla\cdot\left(\left[\textbf{u}-\textbf{u}_{\Gamma}\right] \rho e\right) = -\nabla\cdot\textbf{q} \\
-\nabla\cdot\left(\left[\stress-\mu\rho\identity\right]\cdot\textbf{u}\right)
+ \Sigma\textbf{f}_{ext}\cdot\textbf{u}
\end{split}
\label{eqn:energy}
\end{myequation}
Here, $\rho$ is the density, $\mu=\Sigma\left(\mu_i \rho_i \phi_i\right)/\rho$ the averaged chemical potential per unit mass, $\phi$ the volume fraction of each phase or specie, $\textbf{u}$ the velocity of the fluid, $\textbf{j}_d$ a diffusional mass flux (whose form is irrelevant for the current paper), $\textbf{p}=\rho\textbf{u}$ the linear momentum, $\textbf{f}_{ext}$ the external forces, $\stress = P\identity - \deviatoric$ the stress tensor written in terms of the pressure $P$ and the viscous stress tensor $\deviatoric$, $\rho e$ the total energy per unit volume (internal + kinetic) and $\textbf{q}$ the heat flux. The term corresponding to the chemical potential is added for generality purposes, where variation of the chemical potential produce work (e.g. osmotic pressure). For single phase mixtures assumed to be under chemical equilibrium, the chemical potential and density are constant and, therefore, the contribution of this term to the energy and momentum equations is null. However, for two phase flows, $\nabla\left(\mu\rho\right)$ leads to the surface tension force inside of the phase boundary (e.g. liquid-vapor interface).

An equation for kinetic energy can be obtained by multiplying Eq. (\ref{eqn:momentum}) by the velocity and applying mass conservation:
\begin{myequation}
\begin{split}
\frac{\partial \left(\rho e_{k}\right) }{\partial t}
+\nabla\cdot\left(\left[\textbf{u}-\textbf{u}_{\Gamma}\right] \rho e_{k} +
\left[\stress-\mu\rho\identity \right]\cdot\textbf{u}\right) = \\
\left[\stress-\mu\rho\identity\right]\colon\nabla\textbf{u}
+\Sigma\textbf{f}_{ext}\cdot\textbf{u}
+\frac{\textbf{u}^2}{2}\nabla\cdot\textbf{j}_d
\end{split}
\label{eqn:kinetic}
\end{myequation}
The last term in the RHS of Eq. (\ref{eqn:kinetic}) corresponds to a change in the kinetic energy due to a variation of the density with the local concentration. By subtracting Eq. (\ref{eqn:kinetic}) from Eq. (\ref{eqn:energy}), we obtain an equation for the internal energy:
\begin{myequation}
\begin{split}
\frac{\partial \left(\rho e_{u}\right) }{\partial t}
+\nabla\cdot\left(\left[\textbf{u}-\textbf{u}_{\Gamma}\right] \rho e_{u} \right) = \\
-\nabla\cdot\textbf{q}
-\left[\stress-\mu\rho\identity\right]\colon\nabla\textbf{u}
-\frac{\textbf{u}^2}{2}\nabla\cdot\textbf{j}_d
\end{split}
\label{eqn:internal}
\end{myequation}
Although apparently it does not make much sense to have a term related to kinetic energy in Eq. (\ref{eqn:internal}), its role is to modify the internal energy due to diffusional fluxes changing the concentration. Writing Eq. (\ref{eqn:internal}) in a non-conservatice manner
\begin{myequation}
%\begin{split}
\rho\frac{D^r e_{u}}{D t} =
-\nabla\cdot\textbf{q}
-\left[\stress-\mu\rho\identity\right]\colon\nabla\textbf{u}
+\left(e_{u} -\frac{\textbf{u}^2}{2}\right)\nabla\cdot\textbf{j}_d
%\end{split}
\label{eqn:internal_non_conservative}
\end{myequation}
where $D^r/Dt=\partial/\partial t + \textbf{u}_r\cdot\nabla$ is the material derivative based on the relative velocity between the fluid and the control volume boundary $\textbf{u}_r=\textbf{u}-\textbf{u}_{\Gamma}$. Applying the first law of thermodynamics, the change in the internal energy is given $d E_u = \delta Q - \delta W$, where the heat and work are given by $\delta Q = TdS$ and $\delta W = PdV - \Sigma \mu_i dN_i$ respectively. By dividing the variation of the internal energy by the mass of the system and applying the material derivative leads to
\begin{myequation}
\frac{D^r e_u}{Dt} = T\frac{D^r s_u}{Dt} - P\frac{D^r \rho^{-1}}{Dt}+\Sigma \mu_i\frac{D^r \left(N_i/m\right)}{Dt}
\label{eqn:delta_U}
\end{myequation}
the last term can be re-expressed in terms of an effective chemical potential per unit mass as $\left(\overline{\mu}/\rho \right)D^r\rho/Dt$. Hence, the change in the internal energy is
\begin{myequation}
\frac{D^r e_u}{Dt} = T\frac{D^r s}{Dt} + \frac{1}{\rho^2}\left(P+\overline{\mu}\rho\right)\frac{D^r\rho}{Dt}
\label{eqn:delta_U1}
\end{myequation}
and substituting Eq. (\ref{eqn:delta_U1}) into Eq. (\ref{eqn:internal_non_conservative}) leads to the non-conservation equation for the entropy
\begin{myequation}
\begin{split}
\rho T\frac{D^r s}{Dt} =
-\nabla\cdot\textbf{q}
-\left[\stress-\mu\rho\identity\right]\colon\nabla\textbf{u} \\
-\frac{1}{\rho^2}\left(P+\overline{\mu}\rho\right)\frac{D^r\rho}{Dt}
+\left(e_{u} -\frac{\textbf{u}^2}{2}\right)\nabla\cdot\textbf{j}_d
\end{split}
\label{eqn:entropy}
\end{myequation}
%
%
%For a pure single phase and incompressible system with a fixed control volume, the entropy equation reduces to
%
%
%\begin{equation}
%\begin{split}
%\rho T\frac{D s}{Dt} =
%-\nabla\cdot\textbf{q}
%+\deviatoric\colon\nabla\textbf{u}
%\end{split}
%\label{eqn:entropy2}
%\end{equation}
%
%
%Since the dissipation term changes the entropy, it represents a form internal heat generation. However, and it will be seen soon, this type of heat does not necessarily translate into a change of the enthalpy.
%
By replacing the thermodynamic definition of the enthalpy per unit mass $h=e_u+P\rho^{-1}$ into Eq. (\ref{eqn:internal}), we arrive at
\begin{myequation}
\begin{split}
\frac{\partial \left(\rho h\right) }{\partial t}
+\nabla\cdot\left(\left[\textbf{u}-\textbf{u}_{\Gamma}\right] \rho h + \textbf{q}\right) =
\frac{\partial P }{\partial t}
+\nabla\cdot\left(\left[\textbf{u}-\textbf{u}_{\Gamma}\right] P \right)\\
-\left[\stress-\mu\rho\identity\right]\colon\nabla\textbf{u}
-\frac{\textbf{u}^2}{2}\nabla\cdot\textbf{j}_d
\end{split}
\label{eqn:enthalpy}
\end{myequation}
Considering a pure single phase incompressible system with a fixed control volume ($\textbf{u}_{\Gamma}=0$), the equations for internal energy, entropy and enthalpy respectively reduce to
\begin{myequation}
%\begin{split}
\frac{\partial \left(\rho e_{u}\right) }{\partial t}
+\nabla\cdot\left(\textbf{u}\rho e_{u} \right) =
-\nabla\cdot\textbf{q}+\deviatoric\colon\nabla\textbf{u}
%\end{split}
\label{eqn:internal2}
\end{myequation}
\begin{myequation}
%\begin{split}
\rho T\frac{D s}{Dt} =
-\nabla\cdot\textbf{q}
+\deviatoric\colon\nabla\textbf{u}
%\end{split}
\label{eqn:entropy2}
\end{myequation}
\begin{myequation}
%\begin{split}
\frac{\partial \left(\rho h\right) }{\partial t}
+\nabla\cdot\left(\textbf{u}\rho h + \textbf{q}\right) =
\frac{\partial P }{\partial t}
+\nabla\cdot\left(\textbf{u}P \right)+\deviatoric\colon\nabla\textbf{u}
%\end{split}
\label{eqn:enthalpy2}
\end{myequation}
Equation (\ref{eqn:enthalpy2}) is one of the central results of the present article. Its relevance will be better understood when analyzing the pressure drop for laminar flow in a cylindrical pipe.

\section{Pressure drop for fully developed laminar and turbulent pipe flow}
We start our derivation by integrating Eq. (A15) %in the main text %(\ref{eqn:enthalpy2}) 
over the pipe volume. Since the time derivatives vanish for steady state laminar flows, we have
\begin{myequation}
%\begin{split}
\int_A\rho h \textbf{u}\cdot d\textbf{A} + \int_A\textbf{q}\cdot d\textbf{A} = 
\int_A P\textbf{u}\cdot d\textbf{A} +\int_{\Omega}\deviatoric\colon\nabla\textbf{u}d\Omega 
%\end{split}
\label{eqn:enthalpy3}
\end{myequation}
where the time derivatives vanish due to the steady state condition considered here. By assuming an adiabatic pipe, the surface integral of the heat flux also vanishes. Since for fully developed pipe flow the velocity field has only one component in the axial direction $\textbf{u}=\left(0,0,u_z\right)$, then the tensor velocity gradient in cylindrical coordinates is given by
\begin{myequation}
\nabla \textbf{u} =
\left(
\begin{array}{ccc}
%\frac{\partial u_r}{\partial r} & \frac{1}{r}\frac{\partial u_r}{\partial \varphi} - \frac{u_{\varphi}}{r} & \frac{\partial u_r}{\partial z} \\
%\frac{\partial u_{\varphi}}{\partial r} & \frac{1}{r}\frac{\partial u_{\varphi}}{\partial \varphi} + \frac{u_r}{r} & \frac{\partial u_{\varphi}}{\partial z} \\
%\frac{\partial u_z}{\partial r} & \frac{1}{r}\frac{\partial u_z}{\partial \varphi} + \frac{u_z}{r} & \frac{\partial u_z}{\partial z}\\
0 & 0 & 0 \\
0 & 0 & 0 \\
\frac{\partial u_z}{\partial r} & 0 & 0\\

\end{array}
\right)
\end{myequation}
Thus, the local dissipation is $\deviatoric\colon\nabla\textbf{u}=\mu\left(\frac{\partial u_z}{\partial r}\right)^2 $. Under these conditions, a pressure gradient can only exist along the axial direction and, therefore, the pressure is constant in the pipe cross-section. Hence, Eq. (\ref{eqn:enthalpy3}) reduces to
\begin{myequation}
%\begin{split}
\int_A\rho h \textbf{u}\cdot d\textbf{A} = 
\left(P_{out}-P_{in}\right)VA +\int_{\Omega} \mu\left(\frac{\partial u_z}{\partial r}\right)^2d\Omega 
%\end{split}
\label{eqn:enthalpy4}
\end{myequation}
where $VA=\int_A \textbf{u}\cdot d\textbf{A}$ is the volumetric flow rate. Replacing the analytical solution for parabolic flow $u_z\left(r\right) = 2V\left(1-\left(r/R\right)^2\right)$ into Eq. (\ref{eqn:enthalpy4}) leads to
\begin{myequation}
%\begin{split}
\int_A\rho h \textbf{u}\cdot d\textbf{A} = 
\left(P_{out}-P_{in}\right)VA +8\pi\mu V^2 L
%\end{split}
\label{eqn:enthalpy5}
\end{myequation}     
By integrating independently the momentum equation, we obtain the pressure drop:
$\left(P_{out}-P_{in}\right)VA=-8\pi\mu V^2 L$. This implies that $\int_A\rho h \textbf{u}\cdot d\textbf{A} = 0$, which necessarily requires a constant fluid temperature along the entire pipe. For this particular example, the pressure drop---calculated by integrating the momentum equation---equates exactly to the total dissipation, however, this equality seems unlikely for turbulent or transitional flows. For parabolic laminar flow, the advection term in the momentum equation vanishes, which is not the case for turbulent pipe flows. For us, the determination of the friction factor from the integration of the stresses at the pipe boundary, is incomplete because we are missing all the dissipative processes in the core of the flow. Thus, we believe that a pressure drop expressed in terms of the total dissipation, instead of the wall shear stress, is the most appropriate way to obtain a friction factor.

To investigate the connection between the pressure drop and viscous dissipation in a turbulent flow, we integrate Eq. (A15) %in the main text %(\ref{eqn:enthalpy2}) 
and apply a time average, that is,
\begin{myequation}
%\begin{split}
\int_A  \left\langle \rho h \textbf{u}\right\rangle
\cdot d\textbf{A} = 
\int_A \left(\left\langle\hat{P} \hat{\textbf{u}}\right\rangle + \overline{P}\overline{\textbf{u}}\right) \cdot d\textbf{A} +\int_{\Omega}\left\langle \deviatoric\colon\nabla\textbf{u}\right\rangle d\Omega 
%\end{split}
\label{eqn:enthalpy6}
\end{myequation}
where $\widehat{(\cdot)}$ indicates a fluctuation, and $\overline{(\cdot)}$ and $\left\langle \cdot \right\rangle$ a time average. Since the velocity is zero at walls, the surface integrals reduce only to the inlet and outlet. For a fully--developed turbulent flow, the magnitude of pressure and velocity fluctuations are translation-invariant along the axial direction and, thus, the integration of $\left\langle \widehat{P}\widehat{\textbf{u}}\right\rangle$ vanishes. For a turbulent fully developed pipe flow, the time-averaged pressure $\overline{P}$ is constant in the pipe cross section, hence
\begin{myequation}
%\begin{split}
\int_A  \left\langle \rho h \textbf{u}\right\rangle
\cdot d\textbf{A} = 
\left(\overline{P}_{out}-\overline{P}_{in}\right) VA +\int_{\Omega}\left\langle \deviatoric\colon\nabla\textbf{u}\right\rangle d\Omega 
%\end{split}
\label{eqn:enthalpy7}
\end{myequation}     
For the laminar case it was proved that the enthalpy of the fluid is constant, but no such mathematical proof can be obtained for the turbulent case. However, and since the pressure drop is the only injection of energy to the system (adiabatic case), we have that \mbox{$\int_{\Omega}\left\langle \deviatoric\colon\nabla\textbf{u}\right\rangle d\Omega \leq \left(\overline{P}_{out}-\overline{P}_{in}\right) VA$}. If the energy introduced into the fluid is not fully dissipated, then there cannot by a statistical steady state. Thus, to have a fully developed turbulent pipe flow, the pressure drop must balance the total dissipation. Hence,
\begin{myequation}
\left\langle \Delta P\right\rangle = -\frac{1}{VA}\int_{\Omega}\left\langle \deviatoric\colon\nabla\textbf{u}\right\rangle d\Omega.
\label{eqn:pressure_drop_methods}
\end{myequation}

An obvious experimental validation of our conclusion seems to be simple, but to the best of our knowledge, there is not a single experiment dedicated to measuring directly a possible viscous heating in pipe flow. Sukanek and Laurance \cite{Sukanek} measured indirectly the viscous ``heating'' by measuring the torque needed to rotate two concentric cylinders with an intervening viscous fluid. Since they used a fluid with a viscosity highly dependent on temperature, they argued that a departure from linearity between the shear stress and strain rate would indicate a temperature increase. Indeed, they observed this departure from linearity, but the measured values for the shear stress were much lower than the theoretical ones. The authors attributed this discrepancy to possible air trapped in between the two cylinders. If that is the case, a small amount of trapped air can significantly change the problem. Our analysis is valid strictly for an incompressible flow, and compressibility effects can arise in two-phase flows \cite{Chung} at much lower speeds than sound speed of the phases composing the two-phase mixture.  

When the pressure terms in Eq. (A15) %in the main text %(\ref{eqn:enthalpy2}) 
are omitted, the only logical conclusion is that the temperature must increase due to viscous heating, but this is a conclusion based on an incomplete derivation of the enthalpy equation. If our derivation of the enthalpy equation is wrong and the pressure terms must indeed be omitted, then based on an equilibrium between the injected energy and the dissipation, the enthalpy change must equal the pressure drop. Thus, the temperature difference between the inlet and outlet should be given by $\Delta T = \Delta P/C_p$. This would mean that the pressure drop is completely converted into sensible heat. Our conclusion could be validated or discredited by simply measuring the pressure drop and temperature at the inlet and outlet of a very well insulated long pipe.

%%%%%%%%%%%%%%%%%%%%%%%%%%%%%%%%%%%%%%%%%%%%%%%%%%%%%%%%%%%
%                Dissipation scale                        %
%%%%%%%%%%%%%%%%%%%%%%%%%%%%%%%%%%%%%%%%%%%%%%%%%%%%%%%%%%%
\section{Determining the dissipation scale}
To determine the dissipation scale $\tilde{\epsilon}$, we can make use of our definition of the pressure drop
in terms of the total dissipation Eq. (\ref{eqn:pressure_drop_methods}). This equation can be written as
\begin{myequation}
f = \frac{2 D_H}{V^3}\frac{1}{\Omega} \int_{\Omega}\overline{\epsilon}_0 d\Omega
   +\frac{2 D_H}{V^3}\frac{1}{\Omega} \int_{\Omega}\epsilon'_0 d\Omega
\end{myequation}  

As discussed in the text, the component associated to $\overline{\epsilon}_0$ vanishes for high Reynolds numbers. Thus, to have a constant friction factor as $\Rey\rightarrow\infty$, the second term must be constant. Defining the dissipation scale as
\begin{myequation}
  \tilde{\epsilon}=\frac{1}{\Omega} \int_{\Omega}\epsilon'_0 d\Omega
\end{myequation}
we arrive at once to
\begin{myequation}
\tilde{\epsilon} = C_{\epsilon}\frac{V^3}{D_H}
\label{eqn:dissipation_scale}
\end{myequation} 
with $C_{\epsilon}$ a dimensionless constant. This dissipation scale is consistent with the upper bound derived by Doering and Constantin \cite{Doering} in shear driven turbulence, which supports our definition for the pressure drop.

%%%%%%%%%%%%%%%%%%%%%%%%%%%%%%%%%%%%%%%%%%%%%%%%%%%%%%%%%%%
%                Dissipation scale                        %
%%%%%%%%%%%%%%%%%%%%%%%%%%%%%%%%%%%%%%%%%%%%%%%%%%%%%%%%%%%
\section{Determining the scaling function $\psi\left(R/\eta,\Rey, r/R\right)$}

Using the phenomenological spectrum, the turbulent component of the fraction factor is written as
\begin{myequation}
f_T=\frac{64}{\Rey} \frac{30A \tilde{\epsilon}^{2/3}\Omega}{8\pi V^2}\int_0^\infty q^{1/3}\exp\left(-g\left(\eta,r/D_H\right) q\right) dq
\label{eqn:fl_friction}
\end{myequation}
Making the variable change $y = g\left(\eta,r/D_H\right) q$, the integral is expressed as
\begin{myequation}
I = g\left(\eta,r/D_H\right)^{-4/3}\int_0^\infty y^{1/3}\exp(-y)dy
\end{myequation}
Evaluating the integral leads to
\begin{myequation}
I = g\left(\eta,r/D_H\right)^{-4/3}\frac{2\pi\sqrt{3}}{9\Gamma\left(2/3\right)}
\end{myequation}
with $\Gamma$ the Gamma function. Making use of the dissipation scale Eq. (\ref{eqn:dissipation_scale}), we can express the turbulent component of the friction factor as
\begin{myequation}
f_T=60 D_H^2 A I \left(\frac{C_{\epsilon}}{D_H}\right)^{2/3}\frac{1}{\Rey^{1/4}}\left(\frac{g ^{-4/3}}{\Rey^{3/4}}\right)
\label{eqn:fl_friction1}
\end{myequation}
Comparing this expression to the one provided by Goldelfeld \cite{Nigel}, we arrive to
\begin{myequation}
G\left(\Rey^{3/4}\left(r/R\right)\right) = 60 D_H^2  A I \left(\frac{C_{\epsilon}}{D_H}\right)^{2/3} \left(\frac{g ^{-4/3}}{\Rey^{3/4}}\right)
\end{myequation}
Introducing the Kolmogorov scale, $\eta=\nu^{3/4}\tilde{\epsilon}^{-1/4}$, we arrive to
\begin{myequation}
G\left(\Rey^{3/4}\left(r/R\right)\right) = \gamma \eta g^{-4/3}
\label{eqn:res1}
\end{myequation}
with $\gamma=60 D_H  A I \left(\frac{C_{\epsilon}}{D_H}\right)^{2/3}C_{\epsilon}^{1/4}$ with units of $length^{1/3}$. We now propose the following form for $G$
\begin{myequation}
G\left(\Rey^{3/4}\left(r/R\right)\right) = A_2\left(1-\phi_2(z)\right) + A_3\phi_2(z)\Rey^{1/4}\left(\frac{r}{R}\right)^{1/3}
\label{eqn:res2}
\end{myequation}
where $z = \Rey^{3/4}(r/R)$. The blending function $\phi_2(z)$ provides a smooth transition between the Blasius and Strickles regimes. Following Bobby and Joseph \cite{Bobby2}, we have chosen
\begin{myequation}
 \phi_2(z)=1-\frac{1}{\left(z/z_0\right)^2}
\end{myequation}
By equating Eqs. (\ref{eqn:res1},\ref{eqn:res2}), we arrive to the finally to
\begin{myequation}
g = \eta\left(\frac{\gamma^{3/4}}{\eta^{1/4}\left[A_2\left(1-\phi_2\right)+A_3\phi_2\Rey^{1/4}(r/R)^{1/3}\right]^{3/4}}\right)
\end{myequation}
Since $\gamma^{3/4}$ has units of $length^{1/4}$, the large parenthesis is dimensionless. Thus, we by choosing $\beta = \left(60 A I C_{\epsilon}^{11/12}\right)^{3/4}$ we identify 
\begin{myequation}
\psi = \frac{\left(2R/\eta\right)^{1/4}}{\left[A_2\left(1-\phi_2\right)+A_3\phi_2\Rey^{1/4}(r/R)^{1/3}\right]^{3/4}}
\end{myequation}

%%%%%%%%%%%%%%%%%%%%%%%%%%%%%%%%%%%%%%%
%    Bibliography         %
%%%%%%%%%%%%%%%%%%%%%%%%%%%%%%%%%%%%%%%

%\bibliography{bibliography0}

\begin{thebibliography}{16}
\expandafter\ifx\csname natexlab\endcsname\relax\def\natexlab#1{#1}\fi
\expandafter\ifx\csname bibnamefont\endcsname\relax
  \def\bibnamefont#1{#1}\fi
\expandafter\ifx\csname bibfnamefont\endcsname\relax
  \def\bibfnamefont#1{#1}\fi
\expandafter\ifx\csname citenamefont\endcsname\relax
  \def\citenamefont#1{#1}\fi
\expandafter\ifx\csname url\endcsname\relax
  \def\url#1{\texttt{#1}}\fi
\expandafter\ifx\csname urlprefix\endcsname\relax\def\urlprefix{URL }\fi
\providecommand{\bibinfo}[2]{#2}
\providecommand{\eprint}[2][]{\url{#2}}

\bibitem[{\citenamefont{Barenghi}(2014)}]{Barenghi}
\bibinfo{author}{\bibfnamefont{C.} \bibnamefont{Barenghi~et. al.}},
  \bibinfo{journal}{Proc. Natl. Acad. Sci. USA} \textbf{\bibinfo{volume}{111}},
  \bibinfo{pages}{4647} (\bibinfo{year}{2014}).

\bibitem[{\citenamefont{Tran}(2010)}]{Tran}
\bibinfo{author}{\bibfnamefont{T.} \bibnamefont{Tran~et. al.}},
  \bibinfo{journal}{Nature Phys.} \textbf{\bibinfo{volume}{6}},
  \bibinfo{pages}{438} (\bibinfo{year}{2010}).

\bibitem[{\citenamefont{Richardson}(1922)}]{Richardson}
\bibinfo{author}{\bibfnamefont{L.~F.} \bibnamefont{Richardson}},
  \emph{\bibinfo{title}{Weather Prediction by Numerical Process}}
  (\bibinfo{publisher}{Cambridge University Press}, \bibinfo{year}{1922}).

\bibitem[{\citenamefont{Kolmogorov}(1941{\natexlab{a}})}]{Kolmogorov1}
\bibinfo{author}{\bibfnamefont{A.}~\bibnamefont{Kolmogorov}},
  \bibinfo{journal}{Dokl. Akad. Nauk. SSSR} \textbf{\bibinfo{volume}{30}},
  \bibinfo{pages}{299} (\bibinfo{year}{1941}{\natexlab{a}}).

\bibitem[{\citenamefont{Kolmogorov}(1941{\natexlab{b}})}]{Kolmogorov2}
\bibinfo{author}{\bibfnamefont{A.}~\bibnamefont{Kolmogorov}},
  \bibinfo{journal}{Dokl. Akad. Nauk. SSSR} \textbf{\bibinfo{volume}{32}},
  \bibinfo{pages}{19} (\bibinfo{year}{1941}{\natexlab{b}}).

\bibitem[{\citenamefont{Eyink and Sreenivasan}(2006)}]{Eyink}
\bibinfo{author}{\bibfnamefont{G.}~\bibnamefont{Eyink}} \bibnamefont{and}
  \bibinfo{author}{\bibfnamefont{K.}~\bibnamefont{Sreenivasan}},
  \bibinfo{journal}{Review of Modern Physics} \textbf{\bibinfo{volume}{78}},
  \bibinfo{pages}{87} (\bibinfo{year}{2006}).

\bibitem[{\citenamefont{Maurer and Tabeling}(1998)}]{Maurer}
\bibinfo{author}{\bibfnamefont{J.}~\bibnamefont{Maurer}} \bibnamefont{and}
  \bibinfo{author}{\bibfnamefont{P.}~\bibnamefont{Tabeling}},
  \bibinfo{journal}{Europhysics Letters} \textbf{\bibinfo{volume}{43}},
  \bibinfo{pages}{29} (\bibinfo{year}{1998}).

\bibitem[{\citenamefont{Maltrud and Vallis}(1991)}]{Maltrud}
\bibinfo{author}{\bibfnamefont{M.}~\bibnamefont{Maltrud}} \bibnamefont{and}
  \bibinfo{author}{\bibfnamefont{G.}~\bibnamefont{Vallis}},
  \bibinfo{journal}{J. Fluid Mech.} \textbf{\bibinfo{volume}{228}},
  \bibinfo{pages}{321–341} (\bibinfo{year}{1991}).

\bibitem[{\citenamefont{Smith and Yakhot}(1991)}]{Smith}
\bibinfo{author}{\bibfnamefont{L.}~\bibnamefont{Smith}} \bibnamefont{and}
  \bibinfo{author}{\bibfnamefont{V.}~\bibnamefont{Yakhot}},
  \bibinfo{journal}{J. Fluid Mech.} \textbf{\bibinfo{volume}{228}},
  \bibinfo{pages}{321–341} (\bibinfo{year}{1991}).

\bibitem[{\citenamefont{Navon}(2016)}]{Navon}
\bibinfo{author}{\bibfnamefont{N.} \bibnamefont{Navon~et. al.}},
  \bibinfo{journal}{Nature} \textbf{\bibinfo{volume}{539}},
  \bibinfo{pages}{20114} (\bibinfo{year}{2016}).

\bibitem[{\citenamefont{Gioia and Chakraborty}(2006)}]{Gioia1}
\bibinfo{author}{\bibfnamefont{G.}~\bibnamefont{Gioia}} \bibnamefont{and}
  \bibinfo{author}{\bibfnamefont{P.}~\bibnamefont{Chakraborty}},
  \bibinfo{journal}{Phys. Rev. Lett.} \textbf{\bibinfo{volume}{96}},
  \bibinfo{pages}{044502} (\bibinfo{year}{2006}).

\bibitem[{\citenamefont{Taylor}(1935)}]{Taylor}
\bibinfo{author}{\bibfnamefont{G.}~\bibnamefont{Taylor}},
  \bibinfo{journal}{Proc. Roy. Soc. A} \textbf{\bibinfo{volume}{151}},
  \bibinfo{pages}{421} (\bibinfo{year}{1935}).

\bibitem[{\citenamefont{Goldenfeld}(2006)}]{Nigel}
\bibinfo{author}{\bibfnamefont{N.}~\bibnamefont{Goldenfeld}},
  \bibinfo{journal}{Phys. Rev. Lett.} \textbf{\bibinfo{volume}{96}},
  \bibinfo{pages}{044503} (\bibinfo{year}{2006}).

\bibitem[{\citenamefont{Bobby and Joseph}(2009)}]{Bobby1}
\bibinfo{author}{\bibfnamefont{Y.}~\bibnamefont{Bobby}} \bibnamefont{and}
  \bibinfo{author}{\bibfnamefont{D.}~\bibnamefont{Joseph}},
  \bibinfo{journal}{Journal of Turbulence} \textbf{\bibinfo{volume}{10}},
  \bibinfo{pages}{1} (\bibinfo{year}{2009}).

\bibitem[{\citenamefont{Bobby and Joseph}(2010)}]{Bobby2}
\bibinfo{author}{\bibfnamefont{Y.}~\bibnamefont{Bobby}} \bibnamefont{and}
  \bibinfo{author}{\bibfnamefont{D.}~\bibnamefont{Joseph}},
  \bibinfo{journal}{Physica D} \textbf{\bibinfo{volume}{239}},
  \bibinfo{pages}{1318–1328} (\bibinfo{year}{2010}).

\bibitem[{\citenamefont{Mejia-Alvarez and Christensen}(2010)}]{Mejia}
\bibinfo{author}{\bibfnamefont{R.}~\bibnamefont{Mejia-Alvarez}}
  \bibnamefont{and}
  \bibinfo{author}{\bibfnamefont{K.}~\bibnamefont{Christensen}},
  \bibinfo{journal}{Physics of fluids} \textbf{\bibinfo{volume}{22}},
  \bibinfo{pages}{015106} (\bibinfo{year}{2010}).

\end{thebibliography}

\end{document}